\input harvmac.tex
\input epsf
\noblackbox

\newcount\figno
 \figno=0
 \def\fig#1#2#3{
\par\begingroup\parindent=0pt\leftskip=1cm\rightskip=1cm\parindent=0pt
 \baselineskip=11pt
 \global\advance\figno by 1
 \midinsert
 \epsfxsize=#3
 \centerline{\epsfbox{#2}}
 \vskip 12pt
 {\bf Fig.\ \the\figno: } #1\par
 \endinsert\endgroup\par
 }
 \def\figlabel#1{\xdef#1{\the\figno}}

\def\inbar{\,\vrule height1.5ex width.4pt
depth0pt}

\def\IR{\relax\hbox{$\inbar\kern-.3em{\rm R}$}}
\def\IZ{\relax\hbox{$\inbar\kern-.3em{\rm Z}$}}

\newdimen\tableauside\tableauside=1.0ex
\newdimen\tableaurule\tableaurule=0.4pt
\newdimen\tableaustep
\def\phantomhrule#1{\hbox{\vbox to0pt{\hrule height\tableaurule width#1\vss}}}
\def\phantomvrule#1{\vbox{\hbox to0pt{\vrule width\tableaurule height#1\hss}}}
\def\sqr{\vbox{%
  \phantomhrule\tableaustep
  \hbox{\phantomvrule\tableaustep\kern\tableaustep\phantomvrule\tableaustep}%
  \hbox{\vbox{\phantomhrule\tableauside}\kern-\tableaurule}}}
\def\squares#1{\hbox{\count0=#1\noindent\loop\sqr
  \advance\count0 by-1 \ifnum\count0>0\repeat}}
\def\tableau#1{\vcenter{\offinterlineskip
  \tableaustep=\tableauside\advance\tableaustep by-\tableaurule
  \kern\normallineskip\hbox
    {\kern\normallineskip\vbox
      {\gettableau#1 0 }%
     \kern\normallineskip\kern\tableaurule}%
  \kern\normallineskip\kern\tableaurule}}
\def\gettableau#1 {\ifnum#1=0\let\next=\null\else
  \squares{#1}\let\next=\gettableau\fi\next}

\tableauside=1.0ex \tableaurule=0.4pt


\writedefs

\lref\OSV{OSV}
\newbox\tmpbox\setbox\tmpbox\hbox{\abstractfont }
\Title{\vbox{\baselineskip12pt \hbox{CALT-68-2636}
\hbox{DAMTP-2007-25}\hbox{UT-07-11}
}}
{\vbox{\centerline{Decoupling Supergravity from the Superstring}}}
\vskip 0.2cm

\centerline{Michael B. Green,$^1$ Hirosi Ooguri,$^{2,3}$  and John H. Schwarz$^2$}
\vskip 0.4cm
\centerline{$^1$\it Department of Applied Mathematics and
Theoretical Physics}
\centerline{\it Cambridge University, Cambridge CB3 0WA, UK}
\vskip 0.2cm
\centerline{$^2$\it California Institute of Technology, Pasadena,
CA 91125, USA}
\vskip 0.2cm
\centerline{$^3$\it
Department of Physics, University of Tokyo,
Tokyo 113-0033, Japan}

\vskip 1.3cm

\centerline{\bf Abstract}

We consider the  conditions necessary for obtaining perturbative
maximal supergravity in $d$ dimensions as a decoupling limit of type
II superstring theory compactified on a $(10-d)$-torus.  For
dimensions $d=2$ and $d=3$ it is possible to define a limit in which
the only finite-mass states are the  256 massless states of maximal
supergravity. However, in dimensions $d\ge 4$ there are infinite
towers of additional massless and finite-mass states. These
correspond to Kaluza--Klein charges, wound strings,  Kaluza--Klein
monopoles or  branes wrapping around cycles of the toroidal extra
dimensions. We conclude that perturbative supergravity cannot be
decoupled from string theory in dimensions $\ge 4$. In particular,
we conjecture that pure ${\cal N}=8$ supergravity in four dimensions
is in the Swampland.

\noindent

\bigskip\bigskip
\Date{March, 2007}

\lref\GRV{
  M.~B.~Green, J.~G.~Russo and P.~Vanhove,
  ``Non-renormalisation conditions in type II string theory and maximal
  supergravity,''
  JHEP {\bf 0702}, 099 (2007)
  [arXiv:hep-th/0610299].
}

\lref\BDR{
  Z.~Bern, L.~J.~Dixon and R.~Roiban,
  ``Is ${\cal N} = 8$ supergravity ultraviolet finite?,''
  Phys.\ Lett.\  B {\bf 644}, 265 (2007)
  [arXiv:hep-th/0611086].
}

\lref\BernHH{
  Z.~Bern, J.~J.~Carrasco, L.~J.~Dixon, H.~Johansson, D.~A.~Kosower
and R.~Roiban,
  ``Three-loop superfiniteness of ${\cal N} = 8$ supergravity,''
  arXiv:hep-th/0702112.
}

There has recently has been some speculation that four-dimensional
${\cal N} =8$ supergravity might be ultraviolet finite to all orders
in perturbation theory \refs{\GRV,\BDR,\BernHH}. If true, this would
raise the question of whether ${\cal N} =8$ supergravity might be a
consistent theory that is decoupled from its string theory
extension. A related issue is whether ${\cal N} =8$ supergravity can
be obtained as a well-defined limit of superstring theory. Here we
argue that such a supergravity limit of string theory does not exist
in four or more dimensions, irrespective of whether or not the
perturbative approximation is free of ultraviolet divergences.

\lref\Sen{
  A.~Sen,
  ``D$0$-branes on $T^n$ and matrix theory,''
  Adv.\ Theor.\ Math.\ Phys.\  {\bf 2}, 51 (1998)
  [arXiv:hep-th/9709220].
}

\lref\Seiberg{
  N.~Seiberg,
  ``Why is the matrix model correct?,''
  Phys.\ Rev.\ Lett.\  {\bf 79}, 3577 (1997)
  [arXiv:hep-th/9710009].
}

In this paper, we will study limits of Type IIA superstring theory
on a $(10-d)$-dimensional torus $T^{10-d}$ for various $d$. One may
regard the following analysis  as analogous to the study of the
decoupling limit on D$p$-branes (the limit where field theories on
branes decouple from closed string degrees freedom in the bulk) for
various $p$ \refs{\Sen,\Seiberg}. The decoupling limit on
D$p$-branes is known to exist for $p \leq 5$. On the other hand,
subtleties have been found for $p \geq 6$, where infinitely many new
world-volume degrees of freedom appear in the limit. This has been
regarded as a sign that a field theory decoupled from the bulk does
not exist on D$p$-branes for $p \geq 6$. We will find similar
subtleties for Type IIA theory on $T^{10-d} \times \IR^d$ for $d
\geq 4$.

It will be sufficient for our purposes to consider the torus
$T^{10-d}$ to be the product of $(10-d)$ circles, each of which has
radius $R$. Numerical factors, such as powers of $2\pi$, are
irrelevant to the discussion that follows and therefore will be
dropped. In ten dimensions, Newton's constant is given by
$$
G_{10} =  g^2\ell_{\rm s}^8,
$$
where $\ell_{\rm s}$ is the string scale and $g$ is the string
coupling constant. Thus, the effective Newton constant in $d$
dimensions is given by
\eqn\ndnewton{
G_d \equiv \ell_d^{d-2}
= {G_{10}\over R^{10-d}} = { g^2\ell_{\rm s}^8 \over R^{10-d}},}
where $\ell_d$ is the $d$-dimensional Planck length,
so that
\eqn\coupling{
g = {R^{5-{d\over 2}}\over \ell_{\rm s}^4}\cdot \ell_d^{{d\over 2}-1}.}
We are interested in whether there is a limit of string theory that
reduces to maximal supergravity, which is defined purely in terms of
the dynamics of the 256 states in the massless supermultiplet. In
other words, we are interested in the limit in which all the excited
string states, together with the Kaluza--Klein excitations and
string winding states associated with the $(10-d)$-torus, decouple.
A necessary condition for this to happen is that these states are
all infinitely massive compared to the $d$-dimensional Planck scale
$\ell_d$.\foot{
In this limit, the string length $\ell_{\rm s}$ provides a regularization
scale for supergravity. Thus, if string amplitudes depend sensitively
on $\ell_{\rm s}$, it can be taken as evidence for ultraviolet
divergences in supergravity. This is seen explicitly, for example,
in the one-loop
four graviton amplitude, which is ultraviolet divergent in nine
dimensions. The corresponding string expression is finite and its
low-energy limit is sensitive to the presence of these massive
states with momenta $\sim 1/\ell_{\rm s}$.
}.  This is achieved by taking
\eqn\pertscale{ {1\over R},~~ {1\over \ell_{\rm s}}, ~~{\rm
and}~~{R\over \ell_{\rm s}^2}  \ \gg \ {1\over \ell_d}\,,}
with $\ell_d$ fixed. This is compatible with keeping $g$ fixed for
$d< 6$. If the extra states do decouple then the surviving states
are the 256 massless states of maximal supergravity, which is ${\cal
N}=8$ supergravity when $d=4$.

Let us now consider the spectrum of nonperturbative superstring
excitations in this limit. First consider a D$p$-brane wrapping a
$p$ cycle of the torus. The mass of such a state in $d$ dimensions
is
\eqn\Dbranemass{ M_p = {R^p \over g \ell_{\rm s}^{p+1}}
 = {R^{p+{d\over 2}-5}\over \ell_{\rm s}^{p-3}} \cdot \ell_d^{1-{d \over 2}}.}
When $d \leq 5$, we also need to consider a NS5-brane wrapping a $5$
cycle.  This has a mass given by
\eqn\NSmass{ M_{\rm NS5} = {R^5 \over g^2 \ell_{\rm s}^6}
= {\ell_{\rm s}^2 \over R^{5-d}}\cdot \ell_d^{2-d}.}
In order to obtain the pure supergravity theory with 32 supercharges
in $d$ dimensions, these nonperturbative states also need to
decouple, so their masses must satisfy  $M_p, M_{\rm NS5}\gg
1/\ell_{d}$. In the case of $d=4$ the nonperturbative BPS particle
spectrum also includes Kaluza--Klein monopoles, which are discussed
in the next paragraph.

Before studying the limit in any dimension, $d$, we will discuss
what to expect on general ground. A Kaluza--Klein momentum state
and a wrapped string state have masses $1/R$ and $R/\ell_{\rm s}^2$,
respectively, and they are half-BPS objects that carry a single unit
of a conserved charge. In $d$-dimensions, their magnetic duals are
$(d-4)$-branes. The BPS saturation condition together with the Dirac
quantization condition implies quite generally that the mass $m$ of
a BPS particle and the tension ${\cal T}$ of its magnetic dual
$(d-4)$-brane are related by
\eqn\tension{ m {\cal T} \sim {1\over G_d} = {1 \over \ell_d^{d-2}}.}
Applying this to $d=4$, we immediately conclude that there is no
limit in four dimensions where we can keep all BPS particles heavier
than the Planck scale. In particular, magnetic duals of
Kaluza--Klein excitations, which are the well-known Kaluza--Klein
monopoles, are BPS states with masses $\sim R/\ell_4^2 \rightarrow
0$.\foot{If the torus has six independent radii $R_i$, the
Kaluza--Klein monopole mass spectrum has the form $M^2 =\sum (n_i
R_i / \ell_4^2)^2$.} Similarly, magnetic duals of wrapped strings
are NS5-branes wrapping 5-cycles of $T^6$, and their masses go as
$\ell_{\rm s}^2/R\ell_4^2 \rightarrow 0$. Later, we will discuss
implications of these light states. When $d \ge 5$, at least a
subset of the BPS branes become tensionless in the limit \pertscale.

By contrast, in three dimensions it is possible to define a limit
where all BPS particles become infinitely massive simultaneously. In
this case, magnetic duals of BPS particles are $(-1)$-branes, namely
instantons, and their Euclidean actions vanish in the limit. Thus,
one would expect nonperturbative effects to be very large in three
dimensions even though no singularity is apparent from the spectrum.

In two dimensions, there are no magnetic duals of BPS particles, and
we expect that there is a smooth limit where all BPS particles are
massive and instanton actions remain non-vanishing.

Now, let us look at each case in more detail. When $d=2$, the
conditions we want to impose are
\eqn\twodcondition{ M_p = {1 \over R}
\left({\ell_{\rm s} \over R}\right)^{3-p}~~{\rm and}
~~ M_{\rm NS5} = {1\over R}\left({\ell_{\rm s} \over R}\right)^2
\rightarrow \infty.}
On the other hand, the string coupling constant is given by
\eqn\twocoupling{ g = \left({R \over \ell_{\rm s}}\right)^4.}
Thus, the desired limit can be taken by sending $R\rightarrow 0$
while keeping the string coupling constant finite. In this limit,
all particle masses are much higher than the Planck mass, except for
the massless two-dimensional ${\cal N}=16$ supergravity states
\ref\NicolaiJB{
  H.~Nicolai and N.~P.~Warner,
  ``The structure of ${\cal N}=16$
supergravity in two dimensions,''
  Commun.\ Math.\ Phys.\  {\bf 125}, 369 (1989).
}. However, D$p$-brane and NS5-brane instantons wrapping $T^8$ have
Euclidean actions proportional to $(\ell_{\rm s}/R)^{3-p} \sim
g^{{p-3 \over 8}}$ and $(\ell_{\rm s}/R)^2\sim g^{-{1\over 4}}$,
respectively. Though the actions all remain finite and non-zero in
the limit, their effects are not uniformly suppressed for small $g$.
Thus, the resulting theory may not have a weak coupling limit that
is dominated by the perturbative contribution.

When $d=3$, the conditions we need to impose are
\eqn\threecondition{ M_p = {1 \over \sqrt{R \ell_3}}
\left({\ell_{\rm s} \over R}\right)^{3-p}~~{\rm and}
~~ M_{\rm NS5} = {1\over \ell_3}\left({\ell_{\rm s} \over R}\right)^2
\rightarrow \infty.}
Since we now have
\eqn\threecoupling{ g^2 = {R^7\over \ell_{\rm s}^8} \cdot \ell_3,}
we can rewrite \threecondition\ as
\eqn\masses{ M_p = {g^{{p-3\over 4}} \over R^{{7-p \over
8}}\ell_3^{p+1 \over 8}}~~{\rm and} ~~ M_{\rm NS5} = {1 \over
g^{{1\over 2}} R^{{1\over 4}}\ell_3^{3 \over 4}} \rightarrow
\infty.}
Since $p=0,2,4,6$ in Type IIA theory, this can again be arranged by
taking $R \rightarrow 0$ keeping $g$ finite.\foot{Note that, in the
Type IIB theory, a wrapped D$7$-brane cannot be made heavy unless
$g\gg 1$. This is not in contradiction with T-duality since $g$
transforms under T-duality in such a way that $\ell_{\rm p}$ given
by \ndnewton\ remains invariant. T-duality along one of the circles
on $T^{10-d}$ transforms the coupling $g \rightarrow g \ell_{\rm s}/R$
so it diverges in the limit $R\rightarrow 0$ with the original
coupling constant, given by \threecoupling, kept finite.  } This is
also compatible with the limit \pertscale. Thus, all particle states
develop large masses and may decouple, except for those in
three-dimensional ${\cal N}=16$ supergravity theory \ref\MarcusHB{
  N.~Marcus and J.~H.~Schwarz,
  ``Three-dimensional supergravity theories,''
  Nucl.\ Phys.\  B {\bf 228}, 145 (1983).
}. However, D$p$-brane and NS5-brane instanton actions, which are
given by  $g^{{p-3\over 4}} (R/\ell_3)^{{p+1\over 8}}$ and $g^{-{1\over 2}}
(R/\ell_3)^{{3\over 4}}$, vanish in the limit $R \rightarrow 0$ for any
finite value of $g$. This means that nonperturbative effects are
strong and it may be difficult to determine the properties of the
resulting three-dimensional supergravity.

In view of these observations, it is interesting that gravity
theories formulated in terms of a finite number of fields are known
to exist in two and three dimensions. In three dimensions, the
relation with Chern-Simons gauge theory \ref\WittenHC{
  E.~Witten,
  ``(2+1)-dimensional gravity as an exactly soluble system,''
  Nucl.\ Phys.\  B {\bf 311}, 46 (1988).
} suggests that pure Einstein gravity is finite to all orders in
perturbation theory. However, this theory has no propagating degrees
of freedom, and it is not known whether there is a finite quantum
gravity theory in three dimensions that includes propagating (scalar
or spin-$1/2$) degrees of freedom. Such degrees of freedom are
present, of course, in the examples considered here. The fact that
we find limits of string theory compactifications with a finite
number of such propagating degrees of freedom in these dimensions
may be encouraging, though the implications of the nonperturbative
instanton contributions need to be understood.

When $d=4$, the conditions, \pertscale, necessary for the extra
modes to have infinite masses are
\eqn\fourcondition{ M_p = {1 \over \ell_4}
\left({\ell_{\rm s} \over R}\right)^{3-p}~~{\rm and}
~~ M_{\rm NS5} = {\ell_{\rm s}^2 \over R \ell_4^2}
\rightarrow \infty.}
Clearly, this cannot be realized simultaneously for all $p=0,2,4,6$.
This is in accord with the general argument given earlier, since a
wrapped D$p$-brane and a wrapped D$(6-p)$-brane are
electric--magnetic duals. Similarly, the magnetic duals of
Kaluza--Klein excitations and wrapped strings are Kaluza--Klein
monopoles and wrapped NS5-branes, whose masses behave as
$R/\ell_4^2$ and $\ell_{\rm s}^2/R\ell_4^2$, respectively. There are
infinitely many such states since they have arbitrary integer
charges. In the limit $R, \ell_{\rm s}^2/R \rightarrow 0$, there is
no mass gap and the spectrum becomes continuous.

\lref\Hulla{
  C.~M.~Hull and P.~K.~Townsend,
  ``Unity of superstring dualities,''
  Nucl.\ Phys.\  B {\bf 438}, 109 (1995)
  [arXiv:hep-th/9410167].
}   \lref\Hullb{
  C.~M.~Hull,
  ``String dynamics at strong coupling,''
  Nucl.\ Phys.\  B {\bf 468}, 113 (1996)
  [arXiv:hep-th/9512181].
}
To understand the implications of these infinitely many light states, we
note that among the elements of the four-dimensional U-duality group
$E_7(\IZ)$ is the four-dimensional S-duality transformation that
interchanges the 28 types of electric charge with the corresponding
magnetic charges \refs{\Hulla,\Hullb}.
This duality is described by the following
transformations of the moduli,
\eqn\sdual{
{\rm S}:~  R \rightarrow \tilde R ={\ell_4^2\over R}
~~{\rm and} ~~ \ell_{\rm s} \rightarrow
\tilde\ell_{\rm s} = {\ell_4^2\over \ell_{\rm s}}.}
Note that this transformation inverts the radius $R$ in
four-dimensional Planck units (in contrast to T-duality, which
inverts $R$ in string units). Since $g$ is related to $R$ and
$\ell_{\rm s}$ by \coupling, this transformation  acts as the
inversion $g \rightarrow \tilde g=1/g$, which maps BPS states into
each other. For example, a wrapped D$p$-brane is interchanged with a
wrapped D$(6-p)$-brane. Similarly, a Kaluza--Klein excitation is
interchanged with a Kaluza--Klein monopole (whereas T-duality would
relate it to a wrapped F-string). Thus, in the  dual frame in which
the compactification scale $\tilde R \to \infty$, the six-torus is
decompactified.  This explains the continuous spectrum in the limit
\pertscale. The fact that an infinite set of states from the
nonperturbative sector become massless shows that the limit  of
interest does not result in pure ${\cal N}=8$ supergravity in four
dimensions. Rather, it results in 10-dimensional decompactified
string theory with the string coupling constant inverted. This is
true in both the type IIA and type IIB cases. The only way of
avoiding this would be to relax \pertscale, in which case there
would instead be extra finite-mass Kaluza--Klein or winding number
states, which would therefore not decouple.

One may regard our results on the limit of superstring
compactification on $T^{10-d}$ as examples illustrating the
conjectures formulated in \lref\swampone{
  C.~Vafa,
  ``The string landscape and the swampland,''
  arXiv:hep-th/0509212.
}
\lref\swamptwo{
  H.~Ooguri and C.~Vafa,
  ``On the geometry of the string landscape and the swampland,''
  [arXiv:hep-th/0605264].
} \refs{\swampone,\swamptwo}
on the geometry of continuous moduli parameterizing the string
landscape. The conjectures concern consistent quantum gravity
theories with finite Planck scale in four or more dimensions. Among
the conjectures are the statements that, if a theory has continuous
moduli, there are points in the moduli space that are infinitely far
away from each other, and an infinite tower of modes becomes
massless as a point at infinity is approached \swamptwo.
Since the limit
considered in this paper corresponds to a point in the moduli space
of string compactifications at infinite distance from a generic
point in the middle of moduli space, the conjectures
predict than an infinite number of particles become massless in the
limit. For $d=4$, we have found that among such particles are
Kaluza--Klein monopoles, {\it i.e.}, Kaluza--Klein modes on $T^{6}$
in the dual frame in the limit $\tilde R \rightarrow \infty$. On the
other hand, the moduli space of pure ${\cal N}=8$ supergravity also
contains infinite distance points, but it does not take account of
new light particles appear near these points. If the BPS particles
required by string theory were included one would have string theory
and not ${\cal N} =8$ supergravity.\foot{One can imagine an
alternative history in which type II superstring theory and M-theory
were discovered by properly interpreting the BPS solitons of ${\cal
N} =8$ supergravity.} Thus, the conjectures of \swamptwo\ imply that
the ${\cal N}=8$ supergravity is in the Swampland.
Similarly, there are many superstring compactifications with
${\cal N} < 8$ supersymmetry, and discarding stringy states
in these compactifications results in further supergravity theories
in the Swampland.

It is interesting to see how scattering amplitudes behave in the
limit \pertscale. Consider a four-dimensional graviton scattering
amplitude where the graviton momenta are below the four-dimensional
Planck scale. According to \ndnewton\ and \coupling, the
ten-dimensional Planck length, $\ell_{10}$,  is given by
\eqn\tendnewton{ \ell_{10} = g^{{1\over 4}} \ell_{\rm s}
= R^{{3\over 4}}\ell_4^{1\over4}. }
After the S-duality transformation \sdual, the limit $R \rightarrow
0$ turns into $\tilde R \rightarrow \infty$.  Thus, we have $\tilde
\ell_{10}=\tilde R^{{3\over 4}} \ell_4^{1\over 4}\rightarrow \infty$
in ten dimensions. Since $\tilde \ell_{10} \ll \tilde R$, the extra
dimensions decompactify and the theory is effectively
ten-dimensional. Furthermore, if we take this limit keeping the
graviton momenta fixed (in units of the four-dimensional Planck
mass), the scattering process becomes trans-Planckian. Generically,
we expect that it will involve formation and evaporation of virtual
black holes in ten dimensions.

The original motivation of this work was to investigate the relation
between superstring theory and ${\cal N}=8$ supergravity to see, in
particular, under what conditions supergravity might be ultraviolet
finite. What we have found is that in four or more dimensions ($d\ge
4$) there is no limit of compactified superstring theory in which
the stringy effects decouple and only the 256 massless supergravity
fields survive below the four-dimensional Planck scale. This is true
whether or not there are ultraviolet divergences in supergravity
perturbation theory. Of course, there is a well-defined procedure
for extracting UV finite four-dimensional scattering amplitudes from
perturbative string theory. This involves taking $g \rightarrow 0$
first, before taking the limit \pertscale.  However, this procedure
does not keep $\ell_4$ fixed, and therefore it does not correspond
to the limit considered in this paper.

It might be instructive to compare the situation to that of the
conifold limit of Calabi--Yau compactified type II superstring
theory studied by Strominger \ref\StromingerCZ{ A.~Strominger,
  ``Massless black holes and conifolds in string theory,''
  Nucl.\ Phys.\  B {\bf 451}, 96 (1995)
  [arXiv:hep-th/9504090].
}. In that case, certain terms in the low-energy effective theory
that are independent of the string coupling constant $g$, due to the
decoupling of vector and hypermultiplet fields, can be computed in
string perturbation theory. One can estimate the singularity of
these terms using the fact that a brane wrapping a vanishing cycle
describes a nonperturbative BPS particle that becomes massless in
the conifold limit. If one could identify analogous terms in ${\cal
N}=8$ supergravity, one could transform the Feynman diagram
computation in four-dimensional supergravity into a corresponding
computation in ten dimensions, which might give insight into the
question of ultraviolet finiteness.

The situation is qualitatively different in two and three dimensions
($d=2,3$), where all non-supergravity states develop masses larger
than the Planck scale in the limit \pertscale, and therefore they
can decouple. In these cases only the 256 massless supergravity
states survive, and a self-contained quantum gravity theory may well
exist decoupled from string theory.  We have found, however, that in
the $d=3$ case there are instantons with zero action, which give
rise to large nonperturbative contributions.  In the $d=2$ case the
instanton actions do not vanish in the limit \pertscale, but not all
of them are small when $g$ is small. Therefore the amplitudes may
not be dominated by the perturbative contribution in this case, too.

\bigskip

\centerline{\bf Acknowledgments}

We thank Z.~Bern, N.~Dorey, C.~Hull, J.~Russo, N.~Seiberg, A.~Sen,
M.~Shigemori, Y.~Tachikawa, D.~Tong, P.~Vanhove  and E.~Witten for
discussions. H.O. thanks the hospitality of the particle theory group
of the University of Tokyo.

H.O. and J.H.S. are supported in part by the DOE grant
DE-FG03-92-ER40701. The research of H.O. is also supported in part
by the NSF grant OISE-0403366 and by the 21st Century COE Program
at the University of Tokyo.

\listrefs
\end